\documentclass[runningheads]{llncs}
\usepackage[T1]{fontenc}
\usepackage{graphicx}%
\usepackage{multirow}%
\usepackage{amsmath,amssymb,amsfonts}%
\usepackage{mathrsfs}%
\usepackage[title]{appendix}%
\usepackage{xcolor}%
\usepackage{textcomp}%
\usepackage{manyfoot}%
\usepackage{booktabs}%
\usepackage{algorithm}%
\usepackage{algorithmicx}%
\usepackage{algpseudocode}%
\usepackage{listings}
\usepackage{cite}
\usepackage{adjustbox}
\usepackage{siunitx}
\usepackage{subcaption} 
\usepackage{colortbl} 
\usepackage[table]{xcolor}
\usepackage{verbatim}
\usepackage{longtable}

\usepackage{booktabs} 
\usepackage{multirow} 
\usepackage{graphicx} 
\usepackage{amsmath}  
\usepackage{hyperref}

\usepackage{comment}
\usepackage{siunitx}
\usepackage{xurl}%
\usepackage{url}

\usepackage{tikz}
\usetikzlibrary{trees, positioning}
\usepackage{caption}
\usepackage{tabularx}
\usepackage[table]{xcolor}
\usepackage{colortbl}

\newcounter{promptcounter}

\usepackage{listings}

\makeatletter
\newcommand{\printfnsymbol}[1]{%
  \textsuperscript{\@fnsymbol{#1}}%
}
\makeatother

\begin{document}
\title{Semantically Orthogonal Framework for Citation Classification: Disentangling Intent and Content}
\titlerunning{Framework SOFT: Disentangling Citation Intent and Content}
%
\author{%
  Changxu Duan\inst{1}\thanks{The authors contributed equally and are listed alphabetically.}\orcidID{0000-0003-0547-0901} \and
  Zhiyin Tan\inst{2}\printfnsymbol{1}\orcidID{0009-0002-4166-5810}%
}

\institute{%
  Technische Universität Darmstadt, Darmstadt, Germany \,
  \email{duan@linglit.tu-darmstadt.de}
  \and
  L3S Research Center, Leibniz University Hannover, Hannover, Germany \,
  \email{zhiyin.tan@l3s.de}%
}

\authorrunning{Duan and Tan}
\maketitle              
\begin{abstract}
Understanding the role of citations is essential for research assessment and citation-aware digital libraries. However, existing citation classification frameworks often conflate citation intent (why a work is cited) with cited content type (what part is cited), limiting their effectiveness in auto classification due to a dilemma between fine-grained type distinctions and practical classification reliability. 
We introduce \textbf{SOFT}, a \textbf{S}emantically \textbf{O}rthogonal \textbf{F}ramework with \textbf{T}wo dimensions that explicitly separates citation intent from cited content type, drawing inspiration from semantic role theory. 
We systematically re-annotate the ACL-ARC dataset using SOFT and release a cross-disciplinary test set sampled from ACT2. 
Evaluation with both zero-shot and fine-tuned Large Language Models demonstrates that SOFT enables higher agreement between human annotators and LLMs, and supports stronger classification performance and robust cross-domain generalization compared to ACL-ARC and SciCite annotation frameworks.
These results confirm SOFT's value as a clear, reusable annotation standard, improving clarity, consistency, and generalizability for digital libraries and scholarly communication infrastructures. All code and data are publicly available on GitHub \url{https://github.com/zhiyintan/SOFT}.

\end{abstract}
\section{Introduction}

Citations are central to scholarly communication, shaping how knowledge is credited, organized, and reused across research communities. However, traditional citation metrics, such as counts or h-index, offer only coarse measures of scientific impact, failing to capture the nuanced ways in which works are cited, interpreted, or contested. 
To address this gap, citation classification has emerged as a critical task in computational bibliometrics and information science. It moves beyond simple quantitative counts to unlock the meaning within scholarly dialogue by analyzing the textual context to understand why a work was cited. The aim is to distinguish whether a prior work is being used as a foundational method, critiqued for its findings, or merely mentioned as background. This functional understanding provides the technological backbone for a new generation of digital libraries and intelligent research assistants, enabling advanced services such as automated literature reviews \cite{Cohen2006, JHA2016, jurgens-etal-2018-measuring, Anderson2020Citation}, the detection of scientific controversies \cite{Nicholson2021scite}, and impact-aware scientific search~\cite{valenzuela2015identifying, lo-etal-2020-s2orc}.

Recent advances in Natural Language Processing (NLP) have enabled large-scale citation classification to be deployed in digital library infrastructures. Platforms such as Web of Science~\cite{WoS}, Scite~\cite{Nicholson2021scite}, and Semantic Scholar~\cite{SemanticScholar} incorporate functionalities aimed at detecting aspects of citation intent to improve retrieval, mapping, and assessment. By automatically labeling background, method, result, or support, these platforms enhance transparency, accelerate literature review, and surface influential or controversial contributions~\cite{Dardas2023Evaluating}.

Despite growing adoption, current citation classification practices face persistent limitations. Existing annotation frameworks often conflate distinct citation intents and content types and lack consistent semantic definitions. These ambiguities hinder downstream applications such as citation-based retrieval~\cite{bascur_academic_2023} and impact analysis~\cite{Mercier2021}. 

In this paper, we introduce \textbf{SOFT}: a \textbf{S}emantically \textbf{O}rthogonal \textbf{F}ramework with \textbf{T}wo dimensions for citation annotation. SOFT separates what is being cited (\textbf{Cited Content Type}) from why it is being cited (\textbf{Citation Intent}), enabling clearer definitions, improved annotation consistency, and stronger support for LLM-based classification. Unlike prior one-dimensional frameworks, SOFT is grounded in the principles of semantic roles, offering conceptual alignment with how authors structure argumentation by performing actions (intent) on objects (content).

We validate SOFT through a comprehensive re-annotation of the ACL-ARC dataset~\cite{jurgens-etal-2018-measuring} and a cross-domain evaluation on ACT2~\cite{nambanoor-kunnath-etal-2022-act2}, covering 19 scientific fields. Results show that SOFT yields higher inter-model and human-LLM agreement, improves classification accuracy for both zero-shot and fine-tuned models, and generalizes more robustly across domains than existing frameworks. These findings position SOFT as a reusable annotation standard for supporting citation-aware digital libraries and transparent research assessment.
Our contributions are as follows:
\begin{itemize}
    \item We propose SOFT, a two-dimensional, semantically grounded citation annotation framework that explicitly distinguishes the predicative intent of a citation from its cited content type.
    \item We conduct a systematic re-annotation of the ACL-ARC dataset and release a cross-domain test set from ACT2 under the SOFT framework.
    \item We benchmark SOFT against ACL-ARC and SciCite using zero-shot and fine-tuned LLMs, demonstrating gains in agreement, learnability, and generalizability.
\end{itemize}

\section{Related Work}

Research on citation classification spans several dimensions, including the development of annotation frameworks, the design of automated classification models, and the evaluation of their performance across different datasets and scientific domains. 
Section~\ref{sec:exsitingFrameworks} reviews existing citation intent classification frameworks, most of which rely on single-layer labels that conflate the predicative function of the citation with the type of content being referenced. Section~\ref{sec:cicModels} surveys computational approaches for automating citation intent classification, ranging from early feature-based models to recent prompt-driven large language models. Section~\ref{sec:crossDomain} highlights the limited exploration of cross-domain applicability, particularly across different scientific disciplines, within existing framework designs. Finally, Section~\ref{sec:reannotation} discusses related efforts in re-annotating benchmark datasets, which provide the foundation for our re-annotation of ACL-ARC and ACT2 using a two-dimensional framework.

\subsection{Citation Intent Classification Frameworks}
\label{sec:exsitingFrameworks}

The functional purpose behind why an author cites a particular work has been described using various terms, including \textit{Citation Motivation}~\cite{Spiegel1977Science, Teufel1999ArgumentativeZ}, \textit{Citation Purpose}~\cite{abu-jbara-etal-2013-purpose}, \textit{Citation Type}~\cite{act}, \textit{Citation Function}~\cite{teufel-etal-2006-automatic, jurgens-etal-2018-measuring}, and \textit{Citation Intent}~\cite{cohan-etal-2019-structural}. For conceptual clarity and consistency, we adopt \textbf{Citation Intent} throughout this paper.

Research on citation classification has evolved from bibliometric traditions to computational frameworks tailored for large-scale modeling. One of the earliest and most fine-grained typologies was proposed by~\cite{Spiegel1977Science}, which defined thirteen citation roles spanning rhetorical, epistemic, and contextual functions. 
The Citation Function Corpus (CFC)~\cite{teufel-etal-2006-automatic} was one of the first computationally oriented citation resources. Drawing on Spiegel's typology~\cite{Spiegel1977Science}, it defined twelve function labels grouped into four superclasses (\textsc{Weakness}, \textsc{Contrast}, \textsc{Positive}, and \textsc{Neutral}).
The Citation Typing Ontology (CiTO)~\cite{Shotton2010CiTO} introduced an ontology-based framework with over thirty citation relations, intended for semantic web interoperability. Although formal and expressive, CiTO remains relatively underused in NLP research.

Other frameworks were designed for domain-specific annotation and model training. The six-label framework (\textsc{Use}, \textsc{Criticizing}, \textsc{Comparison}, \textsc{Substantiating}, \textsc{Basis}, and \textsc{Neutral}) proposed by~\cite{abu-jbara-etal-2013-purpose} emphasized purpose-driven types developed through the annotation of NLP publications. 
ACL-ARC~\cite{jurgens-etal-2018-measuring} defined a six-type framework (\textsc{Background}, \textsc{Comparison or Contrast}, \textsc{Motivation}, \textsc{Uses}, \textsc{Extension}, and \textsc{Future}) oriented toward citation influence modeling. 
SciCite~\cite{cohan-etal-2019-structural} condensed this six-type framework into a three-type framework (\textsc{Background}, \textsc{Method}, and \textsc{ResultComparison}), motivated by the observation that types such as \textsc{Motivation}, \textsc{Extension}, and \textsc{Future} often function as background context, to better support machine reading and research navigation in both computer science and biomedical domains. 

ACT and ACT2~\cite{Pride2020Authoritative, nambanoor-kunnath-etal-2022-act2} extended citation annotation to additional scientific fields using a six-type framework similar to ACL-ARC (replacing \textsc{Comparison or Contrast} with \textsc{ComparesContrasts} and extending it to capture not only differences but also similarities and disagreements between citations).
In addition to intent, ACT and ACT2 also annotated influence using two binary labels (\textsc{Influential} and \textsc{Incidental}) with an added emphasis on author influence and comparability. 

Together, these frameworks have shaped the landscape of citation intent classification, offering valuable insights into scholarly discourse. However, many adopt single-layer label structures that implicitly combine the author's action, the citation's function,  with its semantic argument, the type of content being cited. While effective for some tasks, this design can limit semantic clarity and hinder flexible reuse across diverse domains. We return to these structural challenges in Section~\ref{sec:softFramework}, which motivates the design of a new, semantically disentangled framework.

\subsection{Citation Intent Classification (CIC) Models}
\label{sec:cicModels}
Computational approaches to CIC have evolved significantly. Early models relied on hand-engineered features~\cite{qazvinian-radev-2008-scientific, Meng2017AutomaticCO}, but the release of benchmark datasets like ACL-ARC and SciCite catalyzed a shift towards deep learning. Transformer-based encoders, particularly the domain-specific SciBERT~\cite{beltagy-etal-2019-scibert}, quickly became the standard~\cite{devlin-etal-2019-bert, Roman2021Citation}. More recent advances explore diverse learning paradigms, including graph-based models~\cite{Berrebbi2022GraphCite}, multi-task learning~\cite{Ghosal2023Deep, shui-etal-2024-fine}, and prompt-based methods~\cite{Lahiri2023CitePrompt, Kunnath-2023-PromptingStrategies}. Following the latest trend, recent work has also focused on fine-tuning LLMs for this task~\cite{koloveas2025llmspredictcitationintent}. 

In our experiments, we re-annotated the ACL-ARC dataset using our proposed two-dimensional framework and used it for model training and evaluation. To assess cross-domain generalizability, we additionally re-annotated a sample from the ACT2 dataset. Results from these evaluations are presented in Section~\ref{sec:results}, where we compare the performance of CitePrompt and CitationIntentOpenLLM under our framework.

\subsection{Cross-Domain Generalizability}
\label{sec:crossDomain}

A persistent challenge in CIC is the limited ability of models to generalize across different scientific domains or datasets~\cite{shui-etal-2024-fine, yu2024surveyevaluationoutofdistributiongeneralization, li2024probingoutofdistributiongeneralizationmachine}. 
To rigorously assess true generalization capabilities, models must be subjected to cross-domain testing: training on a source dataset (e.g., ACL-ARC) and evaluating on a distinct target (e.g., ACT2)~\cite{chen-etal-2020-cdevalsumm}.

Cross-domain or cross-dataset evaluation offers a principled framework for testing generalization capacity. In this setting, models are trained on one source dataset and evaluated on an independent target dataset~\cite{chen-etal-2020-cdevalsumm}. This setup is widely used across NLP tasks to assess whether models learn domain-agnostic features or overfit to dataset-specific patterns. For instance, prior work has investigated transfer performance in question answering across biomedical, Wikipedia, and web dataset~\cite{guo-etal-2021-multireqa}, and in relation extraction across distinct subdomains such as news and politics~\cite{bassignana-plank-2022-crossre}. 
Another example is the divide in machine translation Generalizability between resource-rich and low-resource languages~\cite{araabi-etal-2023-joint}.

This lack of robustness is particularly evident even when datasets ostensibly share the same annotation framework. For example, ACT2 adopted the ACL-ARC label set for annotations across multiple disciplines. Despite this shared taxonomy, prior work demonstrated a dramatic drop in performance when models effective on ACL-ARC were applied to ACT2~\cite{Kunnath-2023-PromptingStrategies}. This highlights that surface-level label consistency does not guarantee semantic equivalence across domains or robust model transfer.

We interpret these findings as indicative of deeper structural limitations in existing citation classification frameworks. In particular, many current frameworks conflate citation's function with its semantic arguments (the type of cited content), leading to annotation ambiguities that are further amplified in heterogeneous, multi-domain settings. 
In our experiments, we evaluate this by training models on our re-annotated ACL-ARC dataset and testing their performance on the ACT2 dataset using our revised framework annotations.

\subsection{The Applicability of Re-Annotating Datasets}
\label{sec:reannotation}

The performance of supervised machine learning models is fundamentally tied to the quality of their training data~\cite{datacentered}. However, real-world annotations are often noisy, with errors arising from ambiguous guidelines, subjective interpretations, limitations in annotator expertise, or simple human mistakes~\cite{northcutt2021pervasive, Hettiachchi2021Investigating}. Such errors can undermine model training, compromise evaluation reliability, and reduce the robustness of downstream applications.

Recognizing these limitations, recent work has emphasized systematic identification and correction of label errors in widely used benchmarks. For example, studies of the COCO dataset~\cite{Lin2014Microsoft} have revealed deficiencies in object mask annotations, prompting initiatives such as COCO-ReM~\cite{cocorem} and Sama-COCO~\cite{zimmermann2023benchmarkingbenchmarkreliablemscoco}, both of which introduced revised annotations that improved downstream model performance. Similarly, the CleanCoNLL~\cite{rucker-akbik-2023-cleanconll} project conducted a comprehensive relabeling of the CoNLL-03 Named Entity Recognition dataset~\cite{tjong-kim-sang-de-meulder-2003-introduction}, correcting 7\% of entity labels and demonstrating measurable gains in benchmark reliability.

In parallel, automated and semi-automated approaches have been developed to assist error detection and correction. Model-driven methods such as Confident Learning estimate the joint distribution of observed and true labels to identify likely misannotations~\cite{Northcutt2021Confident}, while domain-specific tools like ObjectLab apply trained object detectors to flag anomalies in annotation quality~\cite{Tkachenko2023ObjectLab}. Recently, LLMs have also been explored as annotation agents~\cite{tan-etal-2024-large, Gilardi2023ChatGPT}, although careful quality control remains essential in such settings.

In this work, we conduct a manual re-annotation of the ACL-ARC dataset~\cite{jurgens-etal-2018-measuring} using our proposed two-dimensional framework. For comparison and internal validation, we also task a set of open-source LLMs with performing the same re-annotation under controlled prompting. To evaluate label quality and framework clarity, we compute agreement between human annotators and LLMs. These experiments allow us to assess the reliability of our framework and explore the potential role of LLMs in future annotation workflows.

\section{SOFT: A Semantically Orthogonal Framework with Two Dimensions}
\label{sec:softFramework}

To address the growing need for semantically robust, interpretable, and transferable citation annotations, we propose a new framework grounded in discourse intent and cited content structure. Before introducing its design, we first examine the limitations of existing classification frameworks that motivate our departure from conventional approaches. These limitations arise from structural entanglement, under-specified functional roles, and ambiguous referential targets, all of which hinder consistent annotation and downstream model performance.

\subsection{Limitations of Existing Citation Classification Frameworks}\label{sec:limitation}

Despite the progress enabled by widely used frameworks such as CiTO~\cite{Shotton2010CiTO}, ACL-ARC~\cite{jurgens-etal-2018-measuring}, SciCite~\cite{cohan-etal-2019-structural}, and ACT/ACT2~\cite{act, nambanoor-kunnath-etal-2022-act2}, several structural limitations persist. These challenges reduce annotation clarity and consistency, hinder model generalization, and complicate the semantic interpretation of citation acts. We identify three interrelated problems:

\subsubsection{(1) Dimensional Entanglement: Conflating Cited Content Type and Citation Intent.}
Many previous citation classification frameworks failed to separate the cited content type, such as method, data, or finding, from citation intent or function, such as use, and instead combined both dimensions into a single label. For example, in CiTO, types like ``uses method in'', ``uses data from'', and ``uses conclusions from'' explicitly bind the act of using with the specific object being used, resulting in a conflation of content type and citation intent. This design leads to an explosion in the number of types while still missing important boundary cases, which increases the cognitive burden for annotators and makes modeling more challenging due to both type proliferation and unclassifiable instances.
SciCite, in contrast, presents a related limitation differently: its type \textsc{Background} information is broadly defined and absorbs all citation acts that do not fit into the method or result types. As a result, many citations with specific functional intent are ultimately subsumed under the background type, which obscures their actual scholarly role.

\subsubsection{(2) Perspective Ambiguity: Authorial Commitment and Discourse Voice.}
Additionally, inconsistent perspective across types introduces further confusion for annotators. For example, in the ACL-ARC framework, types such as \textsc{Background}, \textsc{Motivation}, and \textsc{Future} use the cited work as the subject of the action, while types such as \textsc{Uses}, \textsc{Extension}, and \textsc{Comparison or Contrast} use the citing work as the acting subject. This shift in perspective causes cognitive inconsistency and makes the annotation process more confusing. 

\subsubsection{(3) Lack of a Functionally Grounded Intention Framework.}
Moreover, the underlying citation intent is not always clearly specified. For example, a citation marked as \textsc{Comparison or Contrast} might align findings, methods, or conceptual framing, but whether the comparison serves to justify a design, critique a precedent, or merely provide context depends entirely on the citing author's functional goal. Existing frameworks offer no mechanism for disambiguating these divergent functions, leaving the predicative action of the citation act ambiguous. Without an explicit model separating the author's action from the object of that action, annotation relies on inference and subjective interpretation, which reduces reproducibility and cross-annotator agreement.

\subsection{The SOFT Citation Annotation Framework}

We propose \textbf{SOFT}, a two-dimensional citation annotation framework designed to overcome core limitations of previous citation classification systems. SOFT directly addresses three major problems: (1) entanglement of content type and citation function, (2) inconsistent annotation perspectives, and (3) ambiguous or underspecified citation intent.

First, SOFT explicitly disentangles \textbf{Cited Content Type} (what specific content is referenced from the cited work) from \textbf{Citation Intent} (why the citing work refers to that content), thus resolving the long-standing problem of dimensional entanglement, ensuring each dimension is orthogonal and semantically bounded, greatly reducing ambiguity and cognitive load.

Second, SOFT maintains a unified authorial perspective by consistently modeling the action of the citing work. This approach is grounded in semantic role theory~\cite{fillmore1968case, baker-etal-1998-berkeley-framenet, fillmore2006frame, palmer-etal-2005-proposition}, which decomposes events into predicates (actions) and arguments (participants or entities involved). In SOFT, citation intent always corresponds to the action or communicative function performed by the citing work, such as ``use X from [cited work]'', ``modify X from [cited work]'', or ``evaluate against X from [cited work]'', where X is the specific cited content. These intents, like ``use'' or ``modify'', align with classic event predicates in semantic role theory. Other intents in SOFT, such as ``contextualize using X from [cited work]'', ``signal a gap using X from [cited work]'', ``highlight a limitation in X from [cited work]'', or ``justify a design choice based on X from [cited work]'', capture broader discourse actions in scientific writing. While these latter cases extend beyond simple event predicates, all SOFT intent types are defined to specify what the citing work is doing, ensuring each annotation remains grounded in a clear and consistent authorial perspective.

Third, SOFT provides explicit, semantically precise definitions for all citation intent types, ensuring the underlying authorial purpose is clearly specified in every case. For instance, we rephrase \textsc{Background} (used in ACL-ARC, SciCite, and ACT/ACT2) as \textsc{Contextualize} to mark deliberate authorial moves, and rephrase \textsc{Extension} as \textsc{Modify} to capture not only extension but also reduction, replacement, or novel combination. We replace the citation function \textsc{Motivation}(present in ACL-ARC and ACT/ACT2) with \textsc{SignalGap} and \textsc{HighlightLimitation}, which allow for a fine-grained distinction between identifying a research gap and highlighting a limitation. To further reduce ambiguity, we eliminate the type \textsc{Comparison or Contrast} (in ACL-ARC), which is often confused in practice, and instead introduce \textsc{JustifyDesignChoice} for decision justification and \textsc{EvaluateAgainst} for direct empirical comparison.
 
The SOFT type inventory was empirically derived and iteratively refined using diverse datasets and existing frameworks (including SciCite, ACL-ARC, and CiTO), systematically resolving label ambiguity and mixed semantic scope. As a result, SOFT yields a more interpretable and robust annotation framework, improving both human agreement and model generalizability for tasks such as discourse analysis and citation influence detection. Below, we detail the two core dimensions of the framework.

\subsubsection{Dimension 1: Cited Content Type} captures the ontological status of the contribution being referenced in the cited work. We distinguish three types:

\noindent \textbf{\textsc{(1) PerformedWork}}: The citing work references what the cited work \textit{did} (e.g., experimental process, pipeline design), without isolating specific outcomes or reusable resources. \textit{Example: ``[cited work] developed an NLP pipeline.''}

\noindent \textbf{\textsc{(2) Discovery}}: The citing work references observations, findings, or theoretical conclusions made by the cited work. \textit{Example: ``[cited work] observed that dropout improves stability.''}

\noindent \textbf{\textsc{(3) ProducedResource}}: The citing work references reusable outputs such as datasets, algorithms, models, tools, metrics, standardized settings, etc. \textit{Example: ``We use the parser from [cited work].''}

\noindent These distinctions support fine-grained analysis of how different forms of scientific contribution are referenced, reused, or framed in scholarly discourse.

\begin{figure}
\includegraphics[width=0.95\textwidth]{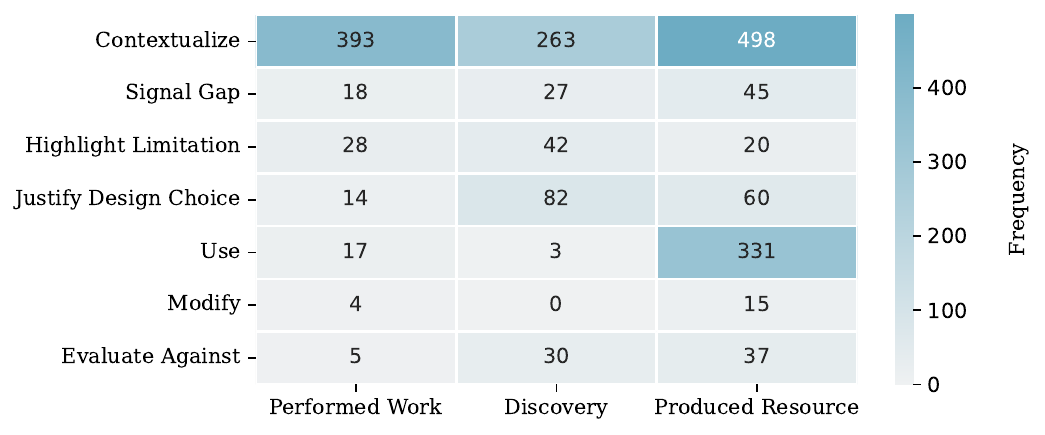}
\caption{Types of statistics for the re-annotated dataset.} \label{fig1}
\end{figure}

\subsubsection{Dimension 2: Citation Intent}
characterizes the predicative function of the citation, representing the action performed by the citing author. 
Each label is defined as a predicate that takes the cited content as one of its core arguments. We frame each label from the perspective of the citing author as the agent of the citation action.

\noindent \textbf{\textsc{(1) Contextualize}}: 
The cited work is mentioned to provide background, illustrate prior research, or describe related contributions. No design decision or reuse is involved. \textit{Example: ``[cited work] introduced hierarchical attention for sentiment analysis.''}

\noindent \textbf{\textsc{(2) SignalGap}}: 
The citation highlights an unresolved problem or open question. The gap may be identified by the cited work or the citing work, and no commitment to solving it is required. \textit{Example: ``[cited work] noted that tuning remains underexplored.''}

\noindent \textbf{\textsc{(3) HighlightLimitation}}: 
The citing work or cited work identifies a flaw, constraint, or drawback in the cited method or result. This type is used only when the cited contribution is explicitly critiqued. \textit{Example: ``[cited work] fails to capture domain drift in user behavior.''}

\noindent \textbf{\textsc{(4) JustifyDesignChoice}}: 
The citing work supports a design or methodological decision by referencing the cited work. No direct reuse is required, but the author must commit to a choice. \textit{Example: ``We follow [cited work]'s feature grouping to ensure consistency.''} This type does not apply if the action is merely hypothetical (e.g., ``we could follow ...'').

\noindent \textbf{\textsc{(5) Use}}: 
The citing work directly applies a reusable contribution (e.g., model, process, settings, definitions) from the cited work. The citing author must be the actor, meaning statements like ``[cited work] used...'' do not qualify. Only past or present-tense application counts; plans or hypothetical uses are not `Use'. \textit{Example: ``We use [cited work]'s BERT-based classifier.''}

\noindent \textbf{\textsc{(6) Modify}}: 
The citing work alters or extends a reusable contribution from the cited work, such as by adapting configurations, changing algorithms, or integrating with a new pipeline. \textit{Example: ``We adapt [cited work]'s encoder for multi-domain inputs.''}

\noindent \textbf{\textsc{(7) EvaluateAgainst}}: 
The citing work explicitly compares its own findings and results with those of the cited work, typically to establish effectiveness. \textit{Example: ``Our model outperforms [cited work]'s method on $F_1$ score.''}

These citation intent types are designed to be mutually exclusive, which can be combined with content types for multi-dimensional interpretation.

\subsubsection{Design Rationale and Applications.}

By structurally separating citation intent from cited content type, SOFT resolves ambiguities common in one-dimensional frameworks. For example, citing a model without applying it is annotated as \textsc{Contextualize:ProducedResource}, while referring to an unresolved issue is \textsc{SignalGap:Discovery}. Only when the citing author explicitly acts, e.g., reuses, modifies, justifies, or critiques, is an active intent assigned. This orthogonality supports consistent labeling across citation contexts and improves annotation reliability.
SOFT’s two-dimensional design also enhances downstream applications. Separating gap signaling from design justification enables clearer modeling of research motivation. Distinguishing between using a tool (\textsc{Use:ProducedResource}) and referencing a methodological trend (\textsc{Contextualize:PerformedWork}) supports finer-grained tracking of scientific influence, method lineage, and intent-aware scientific recommendation. As demonstrated in Section~\ref{sec:results}, the SOFT framework yields higher human–LLM agreement, better classification performance, and improved generalizability across scientific domains.
We next describe our annotation guidelines and procedures, including the re-annotation of ACL-ARC and a cross-domain sample from ACT2.

\section{Experiments}
\subsection{Datasets and Re-Annotation Procedure}
\label{sec:dataset}
\subsubsection{Datasets Overview.}
The ACL-ARC dataset~\cite{jurgens-etal-2018-measuring} is a standard CIC benchmark, yet its annotation framework exhibits the conceptual conflation and boundary issues discussed in Section~\ref{sec:limitation}. To resolve these shortcomings, we re-annotate the entire dataset using our two-dimensional SOFT framework, thereby providing a refined and semantically coherent resource.

For our experiments, we adopt the context-expanded version of ACL-ARC provided by~\cite{nambanoor-kunnath-etal-2022-dynamic}, which enhances the original dataset by adding broader citation contexts and structured metadata, such as the citing and cited paper titles and the section in which the citation appears. Before re-annotation, we manually cleaned the dataset to correct systemic text extraction errors, such as misaligned page breaks and disrupted reading orders, and incorrectly sequenced footnotes. Additionally, we recovered 533 missing section headings from the original paper PDFs to improve contextual understanding.
The ACL-ARC dataset contains 1,931 citation instances. Following the original data split in~\cite{jurgens-etal-2018-measuring}, we adopt the same partitioning: 1,647 instances are used for training and 284 for testing. 

To evaluate cross-domain generalizability, we also re-annotated a subset (264 examples) of the ACT2 test set~\cite{nambanoor-kunnath-etal-2022-act2}. ACT2 spans 19 top-level scientific domains from Microsoft Academic Graph (MAG), with dominant fields including Psychology (21.99\%), Medicine (13.48\%), Biology (10.91\%), and Computer Science (10.27\%). These examples serve as our out-of-distribution test data, enabling strict evaluation of generalization performance.
The division of training and test sets ensures that citation instances from the same source document do not appear in both sets, preventing data leakage and preserving evaluation integrity.

\subsubsection{Re-Annotation Procedure.}
\label{sec:annotation}

The re-annotation process was conducted by a team of graduate students in computational linguistics. To ensure reliability and account for interpretive ambiguity, each citation instance was independently labeled by at least three annotators using our revised two-dimensional framework. Disagreements were flagged and subsequently resolved through structured group discussion, in which annotators reviewed framework definitions and deliberated to reach a consensus type for each case. This multi-pass strategy helped enforce type consistency and refine interpretation standards across annotators.
Types of statistics for the re-annotated ACL-ARC are reported in Figure~\ref{fig1}.

\subsection{Evaluating Framework Interpretability \& Annotation Reliability}

A core indicator of annotation framework quality is Inter-Annotator Agreement (IAA), reflecting how clearly and consistently its types are defined~\cite{artstein-poesio-2008-survey, bayerl-paul-2011-determines}. However, IAA was not reported in the original ACL-ARC release~\cite{jurgens-etal-2018-measuring}, leaving the interpretability of its framework unquantified.
In this study, we also do not report human IAA due to the dynamic annotation process: some annotators departed and new ones joined partway through the task, so not all instances were annotated by the same set of individuals, and the number of annotations per instance varied. This evolving annotator pool is resembles the crowdsourcing approach used for the SciCite dataset~\cite{cohan-etal-2019-structural}, where calculating a dataset-wide human IAA score also presents challenges. Instead, we evaluate framework interpretability and reliability through LLM-based agreement studies.

To assess the clarity and reliability of our framework, we conducted an agreement experiment using open-source LLMs as proxies for human annotators, offering a scalable and replicable assessment of annotation feasibility. Specifically, we deployed four distinct LLMs: \textit{Qwen-2.5-72B-Instruct} (Qwen)~\cite{qwen2025qwen25technicalreport} , \textit{Mistral-Small-24B-Instruct-2501} (Mistral)~\cite{MistralAISmall3_2025}, \textit{Llama-3.3-70B-Instruct} (Llama)~\cite{grattafiori2024llama3herdmodels}, and \textit{Gemma-3-27B-it} (Gemma)~\cite{gemmateam2025gemma3technicalreport}, to independently annotate the ACL-ARC citation contexts according to our framework and guidelines. For each context, every LLM generated a rationale and a classification decision, repeated three times, with the final type determined by majority vote.
We then computed pairwise Cohen's $\kappa$ scores \cite{Cohen1960Coefficient} under two settings: (1) between model pairs (LLM-LLM IAA), and (2) between each model and the human consensus annotations described in Section~\ref{sec:annotation} (Human-LLM IAA). The latter measures how well the framework enables LLMs to approximate expert human judgments.

As a comparative baseline, we repeated the process with both the six-type ACL-ARC framework and the three-type SciCite framework~\cite{cohan-etal-2019-structural}, instructing the same LLMs to annotate ACL-ARC contexts.
For ACL-ARC framework, annotations were made on its original dataset. For SciCite, we mapped ACL-ARC types onto SciCite's framework by aligning \textsc{Uses} with \textsc{Method}, \textsc{ComparisonContrast} with \textsc{ResultComparison}, and grouping \textsc{Background}, \textsc{Extension}, \textsc{Motivation}, and \textsc{Future} under \textsc{Background}.

This LLM-based simulation serves as a practical proxy for human agreement and enables direct empirical comparison across frameworks. The resulting IAA scores provide a quantifiable measure of boundary clarity, interpretability, and annotation feasibility under the SOFT framework.

\subsection{Testing Orthogonality and Interpretive Clarity via Classification}
\label{sec:boundary}
To empirically evaluate whether our revised annotation framework offers clearer semantic boundaries than prior frameworks, we conducted a comparative classification experiment. This design is grounded in the hypothesis, motivated by prior work in framework comparison~\cite{ivanova-etal-2022-comparing, riabi-etal-2025-beyond}, that if a framework defines types with more precise and less ambiguous boundaries, then machine learning models trained on such data should achieve higher classification performance.

\subsubsection{Classifiers.}
We employed two open-source, state-of-the-art CIC models with distinct architectures:
\begin{itemize}
    \item \textbf{CitePrompt}~\cite{Lahiri2023CitePrompt}: A prompt-based fine-tuning framework built on SciBERT~\cite{beltagy-etal-2019-scibert}, implemented via the OpenPrompt toolkit~\cite{ding-etal-2022-openprompt}.
    \item \textbf{CitationIntentOpenLLM}~\cite{koloveas2025llmspredictcitationintent}: A LLM classifier based on \textbf{Qwen-2.5-14B-Instruct} (Qwen-Small), fine-tuned using LLaMA-Factory toolkit~\cite{zheng-etal-2024-llamafactory}.
\end{itemize}
In addition to these fine-tuned models, we also evaluated the performance of several zero-shot classifiers using the LLMs introduced in Section~\ref{sec:annotation}.

The selection of classifiers in this study was guided by a deliberate and strong preference for open-source models over closed-source alternatives. This commitment is crucial for several reasons fundamental to rigorous scientific inquiry. Firstly, utilizing publicly available models like SciBERT and various instruction-tuned LLMs ensures the transparency and reproducibility of our research. This allows other researchers to independently verify our findings and build upon our work without the ``black box'' nature or access restrictions often associated with proprietary systems. Secondly, open-source models grant the critical flexibility required for in-depth analysis and custom fine-tuning, as demonstrated by our use of LLaMA-Factory for adapting Qwen-Small. Such granular control over model architecture and training processes, often unavailable with closed-source APIs, is paramount for rigorously evaluating new annotation frameworks like SOFT and understanding nuanced model behavior. Finally, relying on open-source solutions promotes broader accessibility and fosters a collaborative research environment, which is essential for advancing the field of computational bibliometrics, unhindered by commercial licensing or opaque model mechanics.

\subsubsection{Evaluation Metrics.}

Following prior CIC evaluation protocols~\cite{Lahiri2023CitePrompt, koloveas2025llmspredictcitationintent}, we report both \textbf{Accuracy} and \textbf{Macro-averaged $F_1$ score}. While Accuracy provides an overall measure of correct predictions, Macro $F_1$ computes the average $F_1$ score across classes, mitigating the effects of label imbalance and better reflecting type-level learnability.

\subsection{Evaluating Cross-Domain Generalizability on ACT2}

To assess whether our revised framework supports generalizable citation intent classification across disciplinary contexts, we conducted a transfer evaluation using the test subset of ACT2~\cite{nambanoor-kunnath-etal-2022-act2}, a multi-domain benchmark that spans biology, medicine, economics, and computer science.
As discussed in Section~\ref{sec:exsitingFrameworks}, ACT2 adopts a flat framework derived from ACL-ARC. To adapt it for our evaluation, we selected a representative subset of ACT2 as cross-domain test dataset and applied the same re-annotation protocol described in Section~\ref{sec:annotation} using our two-dimensional framework.
For a controlled comparison, we reused the same classification models (CitePrompt, CitationIntentOpenLLM, and zero-shot LLMs) and evaluation metrics (Accuracy and Macro $F_1$) from Section~\ref{sec:boundary}. This alignment allows us to directly assess whether our framework enhances cross-domain robustness and preserves semantic clarity under domain shift.

\section{Results and Analysis}
\label{sec:results}
We evaluate the SOFT framework along three axes: interpretability, in-domain learnability, and cross-domain generalizability. Results are compared against two existing citation frameworks, ACL-ARC and SciCite.

\definecolor{humanAgree}{RGB}{225, 240, 193}
\begin{table}[htbp] 
\centering 
\caption{Pairwise Cohen's $\kappa$ between LLMs and human annotators.}

\label{tab:agreement_scores_2x2}

\begin{minipage}{0.48\textwidth}
\centering
\subcaption{ACL-ARC Framework (6 types)}\label{tab:agree:acl_arc} 
\begin{tabular}{@{} l 
                   S[table-format=1.4] 
                   S[table-format=1.4]
                   S[table-format=1.4]
                   S[table-format=1.4] @{}} 
\toprule
       & {Llama} & {Mistral} & {Gemma} & {Qwen} \\ 
\midrule
Mistral&  0.5833 &           &         &        \\
Gemma  &  0.4504 &  0.3860   &         &        \\
Qwen   &  0.5776 &  0.6013   &  0.3600 &        \\ 
HUMAN  &  0.3956 &  0.4048   &  0.4302 & 0.3809 \\
\bottomrule
\end{tabular}
\end{minipage}
\hfill 
\begin{minipage}{0.48\textwidth}
\centering
\subcaption{SciCite Framework (3 types)}\label{tab:agree:scicite}
\begin{tabular}{@{} l S[table-format=1.4] S[table-format=1.4] S[table-format=1.4] S[table-format=1.4] @{}}
\toprule
       & {Llama} & {Mistral} & {Gemma} & {Qwen} \\
\midrule
Mistral&  0.6363 &           &         &        \\
Gemma  &  0.6855 &  0.6423   &         &        \\
Qwen   &  0.6762 &  0.7151   &  0.6928 &        \\
HUMAN  &  0.4016 &  0.3850   &  0.4618 & 0.4116 \\
\bottomrule
\end{tabular}
\end{minipage}

\begin{minipage}{0.48\textwidth}
\centering
\subcaption{SOFT: Cited Content Type (3 types)}\label{tab:agree:object}
\begin{tabular}{@{} l S[table-format=1.4] S[table-format=1.4] S[table-format=1.4] S[table-format=1.4] @{}}
\toprule
       & {Llama} & {Mistral} & {Gemma} & {Qwen} \\
\midrule
Mistral&  0.5396 &           &         &        \\
Gemma  &  0.5751 &  0.5657   &         &        \\
Qwen   &  0.5496 &  0.6169   &  0.5717 &        \\
HUMAN  &  \cellcolor{humanAgree}0.5193 &  0.4932   &  \cellcolor{humanAgree}\textbf{0.5901} & \cellcolor{humanAgree}0.5596 \\
\bottomrule
\end{tabular}
\end{minipage}
\hfill
\begin{minipage}{0.48\textwidth}
\centering
\subcaption{SOFT: Citation Intent (7 types)}\label{tab:agree:function}
\begin{tabular}{@{} l S[table-format=1.4] S[table-format=1.4] S[table-format=1.4] S[table-format=1.4] @{}}
\toprule
       & {Llama} & {Mistral} & {Gemma} & {Qwen} \\
\midrule
Mistral&  0.6154 &           &         &        \\
Gemma  &  0.5941 &  0.5949   &         &        \\
Qwen   &  0.6282 &  0.5887   &  0.6175 &        \\
HUMAN  &  \cellcolor{humanAgree}0.6620 &  \cellcolor{humanAgree}0.6070   &  \cellcolor{humanAgree}0.6232 & \cellcolor{humanAgree}\textbf{0.6918} \\
\bottomrule
\end{tabular}
\end{minipage}
\end{table}

\subsubsection{Interpretability and Agreement.}
We assess framework clarity using pairwise Cohen's $\kappa$ across four pre-trained LLMs (LLM-LLM) and between LLM predictions and human consensus labels (Human-LLM). As expected, SciCite achieves the highest LLM-LLM agreement due to its coarse granularity. However, as shown in Table~\ref{tab:agreement_scores_2x2}, SOFT offers significantly higher Human-LLM agreement, especially for the \textbf{Citation Intent} dimension ($\kappa=0.662$ for Llama, $0.692$ for Qwen). Despite being conceptually more complex, \textbf{Cited Content Type} also outperforms ACL-ARC and SciCite in Human-LLM agreement, indicating that both dimensions are interpretable and consistently applicable.

\subsubsection{In-Domain Learnability.}
We evaluate the classification performance of both fine-tuned and zero-shot models on the ACL-ARC test set using three annotation frameworks: ACL-ARC, SciCite, and SOFT. In the SOFT framework, \textbf{Cited Content Type} and \textbf{Citation Intent} are treated as two independent classification tasks. Table~\ref{tab:model_performance} and Figure~\ref{fig:grid_radar_plots} in Appendix~\ref{appendix:perTypePerformance} report accuracy and macro $F_1$ scores across all models.
Despite its conceptual specificity, \textbf{Cited Content Type} classification yields strong in-domain performance. Fine-tuned Qwen-Small achieves 0.78 macro $F_1$, while zero-shot models average 0.66. Although no existing framework provides a comparable baseline, these results demonstrate that content type distinctions are not only human-interpretable but also model-learnable across architectures.
SOFT also outperforms prior frameworks on \textbf{Citation Intent} classification. For example, fine-tuned Qwen-Small achieves 0.65 macro $F_1$ under SOFT, compared to 0.51 on ACL-ARC and 0.62 on SciCite. Notably, this gain occurs even though SOFT-Intent has more types (7) than ACL-ARC (6), refuting the assumption that performance is merely a function of label count. This suggests the framework's semantic clarity and well-defined boundaries are more critical to model performance than raw granularity. The average zero-shot $F_1$ under SOFT reinforces this point, reaching 0.69, substantially above ACL-ARC (0.49) and SciCite (0.61), demonstrating that SOFT's functional types align more closely with both pre-trained and fine-tuned model representations.

We note that our reported scores are lower than those in prior work, including CitePrompt~\cite{Lahiri2023CitePrompt}, which may be attributed to differences in dataset partitioning, as discussed in Section~\ref{sec:dataset}.

\subsubsection{Cross-Domain Generalizability.}
\label{sec:crossdomain}
\begin{figure}
\includegraphics[width=0.9\textwidth]{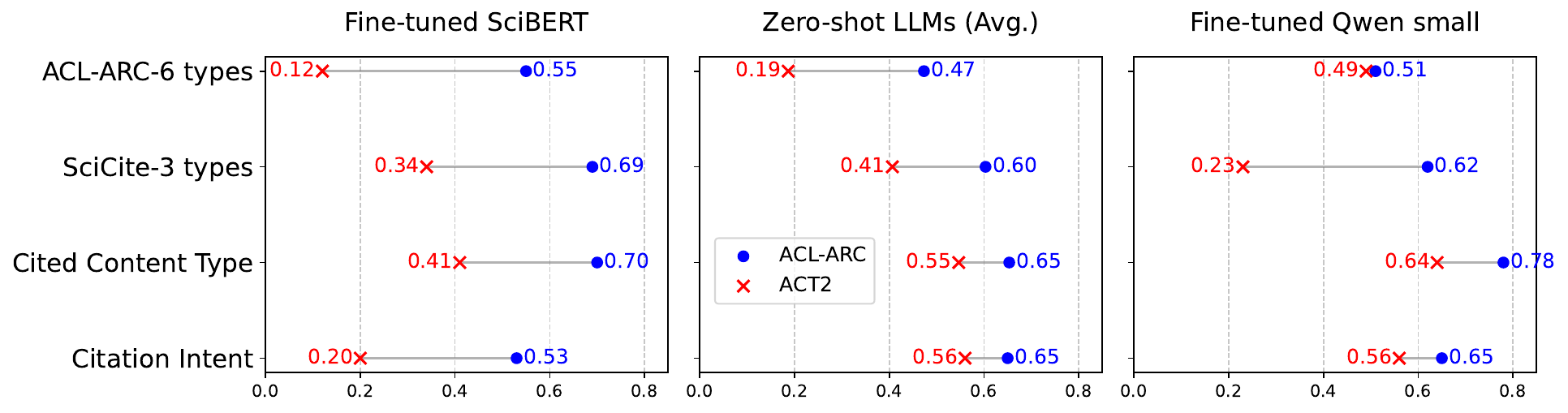}
\caption{Comparison of in-domain and cross-domain macro $F_1$ scores across four annotation frameworks. Each subplot corresponds to a model type, with blue circles denoting in-domain and red crosses indicating cross-domain performance.} \label{fig:drops}
\end{figure}

To assess the robustness of different frameworks under domain shift, we evaluate models trained on ACL-ARC directly on a re-annotated subset of ACT2. No further fine-tuning or domain adaptation is performed. Table~\ref{tab:model_performance} and Figure~\ref{fig:grid_radar_plots} in Appendix~\ref{appendix:perTypePerformance} report both accuracy and macro $F_1$ scores, highlighting how well each framework generalizes beyond the original training domain.
Despite its conceptual complexity, \textbf{Cited Content Type} classification generalizes well. Zero-shot LLMs maintain solid performance (average $F_1$ drop 17\%), and fine-tuned Qwen-Small drops moderately (0.78 to 0.64), suggesting that these content distinctions remain stable across domain shifts. 
SOFT-trained models exhibit superior generalizability on \textbf{Citation Intent} classification. Fine-tuned Qwen-Small drops from 0.65 to 0.56 (SOFT), outperforming ACL-ARC (0.51 to 0.49) and SciCite (0.62 to 0.23).
Zero-shot performance mirrors this trend (Figure~\ref{fig:drops}). 
SOFT's average $F_1$ drop is only 15.8\% (vs. 30-65\% for other frameworks), suggesting its discourse-grounded intent types generalize more robustly.

\subsubsection{Dimension-Wise Comparison.}
A dimension-wise analysis reveals distinct model behaviors. In-domain, zero-shot LLMs perform better on \textbf{Citation Intent} (avg. $F_1$: 0.69) than \textbf{Cited Content Type} (0.66), suggesting these functions align well with LLM priors. 
With fine-tuning, however, \textbf{Cited Content Type} becomes more learnable (Qwen-Small $F_1$: 0.78 vs. 0.65 for intent), indicating its distinctions benefit more from supervision. 
Cross-domain generalization is competitive for both dimensions, with fine-tuned Qwen showing only modest performance drops on both content (Qwen-Small: 18\%; SciBERT: 41\%) and intent (Qwen-Small: 14\%; SciBERT: 62\%). 
This highlights SOFT's stability under domain shift and the superior transferability of fine-tuned Qwen over SciBERT.

\subsubsection{Overall Analysis.}
Across all experiments, the SOFT framework consistently outperforms ACL-ARC and SciCite in interpretability, learnability, and cross-domain robustness. Fine-tuned Qwen-Small achieves the best results on \textbf{Cited Content Type}, while Qwen zero-shot achieves top performance on \textbf{Citation Intent}. These findings establish SOFT as a semantically orthogonal and generalizable alternative for citation annotation.

\section{Conclusion}
This paper introduces the SOFT framework, a semantically orthogonal framework with two dimensions for citation annotation that disentangles \textit{cited content type} from \textit{citation intent}. By addressing the conceptual conflation and boundary ambiguity in existing frameworks, SOFT provides clearer interpretive definitions and better supports both human annotation and model-based classification.  Empirical evaluations across LLM agreement, in-domain classification, and cross-domain generalization confirm SOFT's advantages. The \textit{citation intent} dimension demonstrates high zero-shot and fine-tuned performance, while the \textit{cited content type} dimension, despite its semantic complexity, proves learnable and transferable. Compared to ACL-ARC and SciCite, SOFT consistently achieves higher human-LLM agreement, stronger in-domain macro $F_1$, and more stable cross-domain generalization. These gains are especially pronounced for instruction-tuned LLMs, which preserve broader generalization after fine-tuning.  

Future work will explore extending SOFT to support multi-label citation contexts, integrating finer-grained functional types, and applying the framework to large-scale citation graphs for discourse-level functional modeling, influence tracking, and knowledge extraction. 

\section*{Acknowledgment}
Changxu Duan was funded by the Federal Ministry of Education and Research (BMBF) under grant no. 01UG2130A, as part of the InsightsNet research project (\url{insightsnet.org}). 
Zhiyin Tan was funded by the ``HybrInt - Hybrid Intelligence through Interpretable AI in Machine Perception and Interaction'' project (Zukunft Nds, Niedersächsisches Ministerium für Wissenschaft, Grant ID: ZN4219).
We thank Dr. Sabine Bartsch for proofreading.
We also thank the anonymous reviewers for their insightful comments and suggestions.

%
%
%
\bibliographystyle{splncs04}
\bibliography{custom}

\appendix

\section{Overall Model Performance}
\label{appendix:overallPerformance}

Table \ref{tab:model_performance} presents the overall performance of the evaluated models across the ACL-ARC, SciCite, and our proposed SOFT frameworks, for both in-domain (ACL-ARC test set) and cross-domain (re-annotated ACT2 test set) scenarios. The performance is measured by accuracy and macro-$F_1$ score.

\section{Per-Type Analysis Across Frameworks and Models}
\label{appendix:perTypePerformance}

\begin{figure}[htbp]
    \centering
    \includegraphics[width=\textwidth]{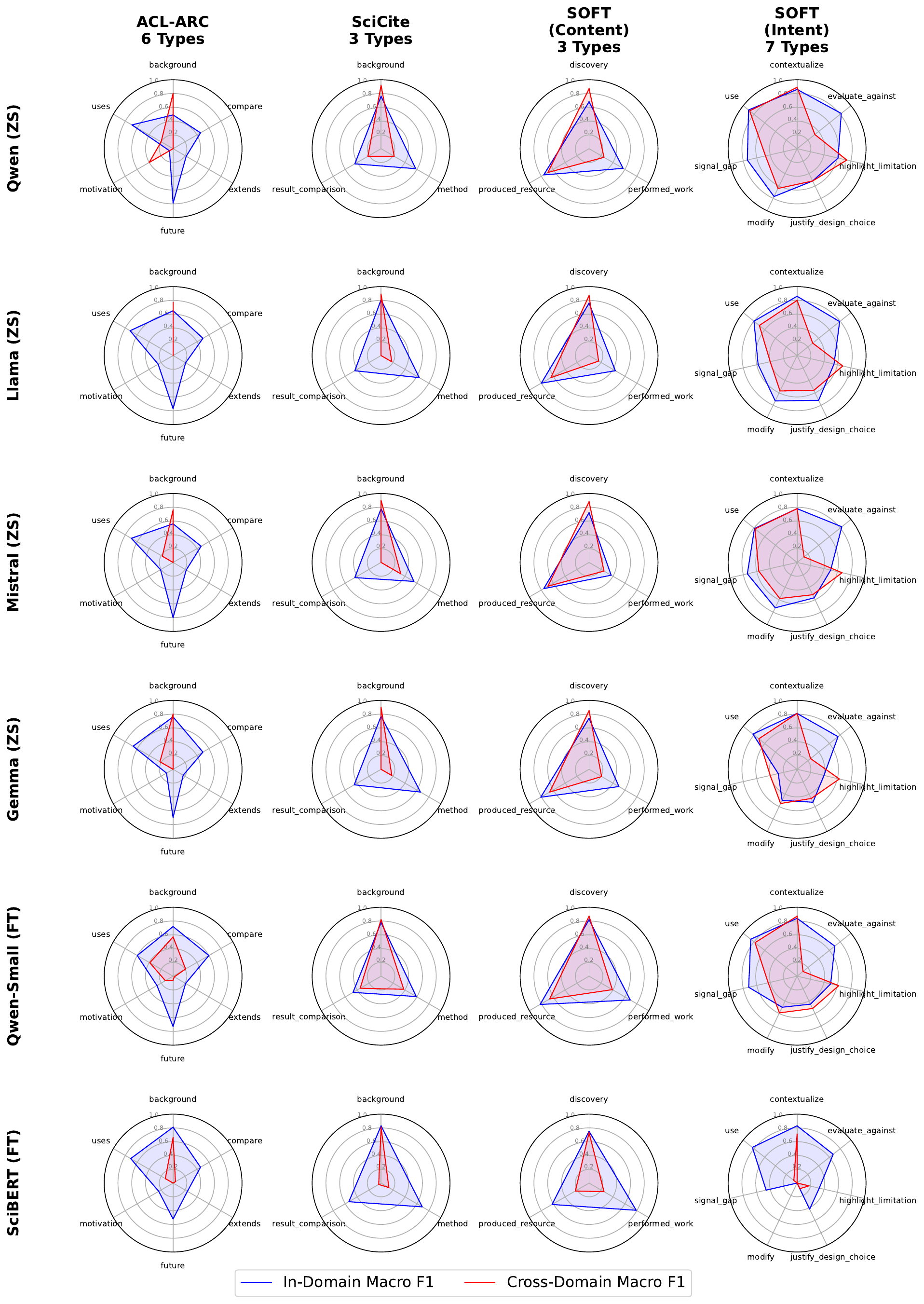}
    \caption{Grid of radar plots (4 frameworks $\times$ 6 models) showing per-class F1-scores.
    Blue lines/areas represent In-Domain F1 scores, and red lines/areas represent Cross-Domain F1 scores.
    Each axis corresponds to a class label within the respective framework. F1-scores range from 0 (center) to 1 (outer edge).}
    \label{fig:grid_radar_plots}
\end{figure}

To gain a more granular understanding of model behavior and framework characteristics, we visualize the per-class $F_1$-scores for both in-domain and cross-domain settings using radar plots, as shown in Figure \ref{fig:grid_radar_plots}. 
Each plot depicts the performance of a specific model across the classes of a given framework, with blue representing in-domain and red representing cross-domain $ F_1$ Scores. These visualizations allow for a direct comparison of how well individual classes are learned and how performance on these classes transfers across domains.

Across ACL-ARC and SciCite, only a few types achieve consistently learnable performance. In ACL-ARC, \textsc{Uses} ($F_1$ = 0.60-0.72) and \textsc{Future} ($F_1$ = 0.70-0.80) are reliably detected across models, while types like \textsc{Motivation} ($F_1$ = 0.06-0.27) and extends ($F_1$ = 0.17-0.24) exhibit uniformly poor results, regardless of model type. In SciCite, background and method achieve moderate $F_1$ (up to 0.83 and 0.69), but \textsc{ResultComparison} remains consistently weak ($F_1\leq0.47$, recall < 0.5). These patterns suggest semantic overlap and label fuzziness in legacy frameworks, especially for tasks that require disambiguating subtle intent.

The SOFT framework, in contrast, features multiple types that are both conceptually precise and computationally learnable. Zero-shot LLMs perform well on \textsc{Use} ($F_1$ = 0.79-0.90), \textsc{Contextualize} (0.78-0.86), and \textsc{ProducedResource} (0.76-0.82), while the fine-tuned LLM (Qwen-small) outperforms all other models on discovery and \textsc{ProducedResource} (both $F_1$ = 0.82). More difficult types like \textsc{PerformedWork} show a large performance gap in $F_1$ score (LLMs: 0.37-0.57, Qwen-small: 0.69, SciBERT: 0.79), indicating that domain-adaptive fine-tuning is crucial when lexical variation is high. For \textsc{JustifyDesignChoice}, zero-shot LLM Llama ($F_1 = 0.72$) outperforms both fine-tuned models (Qwen-small: $F_1$ = 0.45, SciBERT: $F_1$ = 0.42), suggesting that general discourse modeling, rather than label-specific exposure, governs success.

Precision-recall asymmetries further differentiate framework quality. In SciCite, the method exhibits high recall (e.g., zero-shot LLMs: 0.86-0.94) but low precision (e.g., zero-shot LLMs: 0.4-0.53), indicating label overgeneralization. In SOFT, several types, including \textsc{JustifyDesignChoice} and performed work, exhibit the opposite: high precision (e.g., Qwen: 0.79) but low recall (e.g., Mistral: 0.28), showing that models can detect prototypical cases but struggle with broader coverage. Notably, \textsc{SignalGap} (Qwen: precision 0.81, recall 0.68) and \textsc{Method} (SciBERT: precision 0.64, recall 0.76) achieve better balance. These contrasts affirm that our SOFT framework not only yields more distinguishable and learnable labels but also makes model errors more interpretable, enabling both zero-shot application and targeted fine-tuning.

\section{Computational Resources and Software}
We adopted the original hyperparameter settings from CitePrompt~\cite{Lahiri2023CitePrompt} and CitationIntentOpenLLM~\cite{koloveas2025llmspredictcitationintent}. All experiments were conducted on a single H100 GPU, including both fine-tuning and inference. For LLM inference, we used the vLLM library~\cite{kwon2023efficient}. Fine-tuning required approximately 5 GPU-hours, and inference consumed an additional 4 GPU-hours in total.

\definecolor{light}{RGB}{253, 243, 243} 
\definecolor{mid}{RGB}{250, 225, 225} 
\definecolor{dark}{RGB}{244, 189, 193} 

\begin{table}[htbp]
\centering
\caption{In-domain and cross-domain performance of citation classification models (zero-shot: ``ZS'', fine-tuned: ``FT'') across frameworks.}
\label{tab:model_performance}
\begin{tabular}{@{}ll c >{}c c >{}c @{}} 
\toprule
\multirow{2}{*}{\textbf{Framework}} & \multirow{2}{*}{\textbf{Classifier}} & \multicolumn{2}{c}{\textbf{In-Domain (ACL-ARC)}} & \multicolumn{2}{c}{\textbf{Cross-Domain (ACT2)}} \\
\cmidrule(lr){3-4} \cmidrule(lr){5-6} 
 &  & Accuracy & Macro $F_1$ & Accuracy & Macro $F_1$ \\
\midrule

\multirow{6}{*}{\parbox{1.7cm}{\centering ACL-ARC\\(6 types)}} & ZS Llama  & 0.58 & 0.52 & 0.60 & \cellcolor{light}0.13 (75\%↓) \\
                             & ZS Mistral & 0.54 & 0.49 & 0.59 & \cellcolor{light}0.16 (67\%↓) \\
                             & ZS Gemma    & 0.65 & 0.48 & 0.62 & \cellcolor{light}0.17 (65\%↓) \\
                             & ZS Qwen  & 0.48 & 0.45 & \textbf{0.63} & \cellcolor{light}0.23 (49\%↓) \\
                             \cmidrule(lr){2-6}
                             & FT SciBERT & \textbf{0.66} & \textbf{0.55} & 0.47 & \cellcolor{light}0.12 (78\%↓) \\
                             & FT Qwen-Small & 0.65 & 0.51 & 0.58 & \cellcolor{dark}\textbf{0.49} ( 4\%↓) \\
\midrule

\multirow{6}{*}{\parbox{1.7cm}{\centering SciCite\\(3 types)}} & ZS Llama  & 0.72 & 0.63 & 0.80 & \cellcolor{light}0.37 (54\%↓) \\
                                 & ZS Mistral  & 0.67 & 0.59 & 0.81 & \cellcolor{mid}0.41 (31\%↓) \\
                                 & ZS Gemma    & 0.70 & 0.63 & 0.80 & \cellcolor{light}0.36 (43\%↓) \\
                                 & ZS Qwen  & 0.67 & 0.59 & \textbf{0.84} & \cellcolor{mid}\textbf{0.45} (24\%↓) \\
                                 \cmidrule(lr){2-6}
                                 & FT SciBERT & \textbf{0.77} & \textbf{0.69} & 0.66 & \cellcolor{light}0.34 (51\%↓) \\
                                 & FT Qwen-Small & 0.69 & 0.61 & 0.40 & \cellcolor{light}0.23 (63\%↓)\\
\midrule

\multirow{6}{*}{\parbox{1.7cm}{\centering \textbf{SOFT}:\\Cited Content\\Type\\(3 types)}} & ZS Llama    & 0.73 & 0.67 & 0.75 & \cellcolor{dark}0.55 (18\%↓) \\
                            & ZS Mistral  & 0.67 & 0.62 & \textbf{0.78} & \cellcolor{mid}0.46 (26\%↓) \\
                            & ZS Gemma    & 0.74 & 0.68 & 0.73 & \cellcolor{dark}0.58 (15\%↓) \\
                            & ZS Qwen  & 0.70 & 0.67 & 0.77 & \cellcolor{dark}0.60 (10\%↓) \\
                            \cmidrule(lr){2-6}
                            & FT SciBERT & 0.70 & 0.70 & 0.58 & \cellcolor{light}0.41 (41\%↓)\\
                            & FT Qwen-Small & \textbf{0.79} & \textbf{0.78} & 0.75 & \cellcolor{dark}\textbf{0.64} (18\%↓) \\
\midrule

\multirow{6}{*}{\parbox{1.7cm}{\centering \textbf{SOFT}:\\Citation Intent\\(7 types)}} & ZS Llama   & 0.79 & 0.72 & 0.72 & \cellcolor{mid}0.57 (21\%↓)\\
                              & ZS Mistral  & 0.74 & 0.71 & 0.69 & \cellcolor{dark}0.57 (20\%↓) \\
                              & ZS Gemma    & 0.71 & 0.59 & 0.71 & \cellcolor{dark}0.55 ( 7\%↓) \\
                              & ZS Qwen  & \textbf{0.81} & \textbf{0.75} & \textbf{0.82} & \cellcolor{dark}\textbf{0.64} (15\%↓) \\
                              \cmidrule(lr){2-6}
                              & FT SciBERT & 0.72 & 0.53 & 0.52 & \cellcolor{light}0.20 (62\%↓) \\
                              & FT Qwen-Small & 0.77 & 0.65 & 0.77 & \cellcolor{dark}0.56 (14\%↓) \\
\bottomrule
\end{tabular}
\begin{minipage}{\linewidth}
\small
\textit{Color Legend:} 
\colorbox{dark}{$\leq$20\% drop},\,
\colorbox{mid}{21--40\% drop},\,
\colorbox{light}{>40\% drop} in cross-domain Macro $F_1$.
\end{minipage}
\end{table}

\end{document}